\documentclass[floatfix,nofootinbib,twocolumn,showpacs,superscriptaddress, reprintnumbers,amssymb,letterpaper,amsmath,pra]{revtex4-2}

\usepackage[scientific-notation=true]{siunitx}
\usepackage{graphicx}
\usepackage{dcolumn}
\usepackage{bm}
\usepackage{xcolor}
\usepackage[colorlinks=true,allcolors=blue]{hyperref}
\usepackage{amsmath}
\usepackage{lineno}
\usepackage{float}
\usepackage{tablefootnote}
\usepackage{scrextend}
\usepackage{braket}
\usepackage{calrsfs}
\DeclareMathAlphabet{\pazocal}{OMS}{zplm}{m}{n}
\newcommand{\Lb}{\pazocal{L}}
\usepackage{tabstackengine}
\stackMath

\usepackage{array}
\newcolumntype{C}[1]{>{\centering\arraybackslash}p{#1}}

\begin{document}
\title
{
Four-Dimensional Phase-Space Reconstruction of Flat and Magnetized Beams Using Neural Networks and Differentiable Simulations
}

\author{Seongyeol Kim}\email{sykim12@postech.ac.kr}\thanks{now at Pohang Accelerator Laboratory, POSTECH, Republic of Korea.}
\affiliation{Argonne National Laboratory, Lemont, IL 60439, USA}
\author{Juan Pablo Gonzalez-Aguilera}\email{jpga@uchicago.edu}
\affiliation{Department of Physics and Enrico Fermi Institute, University of Chicago, Chicago, IL 60637, USA}
\author{Philippe Piot}
\affiliation{Argonne National Laboratory, Lemont, IL 60439, USA}
\affiliation{Northern Illinois Center for Accelerator \& Detector Development and Department of Physics, Northern Illinois University, DeKalb, IL 60115, USA} 
\author{Gongxiaohui Chen}
\affiliation{Argonne National Laboratory, Lemont, IL 60439, USA}
\author{Scott Doran}
\affiliation{Argonne National Laboratory, Lemont, IL 60439, USA} 
\author{Young-Kee Kim}
\affiliation{Department of Physics and Enrico Fermi Institute, University of Chicago, Chicago, IL 60637, USA}
\author{Wanming Liu}
\affiliation{Argonne National Laboratory, Lemont, IL 60439, USA}
\author{Charles Whiteford}
\affiliation{Argonne National Laboratory, Lemont, IL 60439, USA}
\author{Eric Wisniewski}
\affiliation{Argonne National Laboratory, Lemont, IL 60439, USA}
\author{Auralee Edelen}
\affiliation{SLAC National Accelerator Laboratory, Menlo Park, CA 94025, USA}
\author{Ryan Roussel}
\affiliation{SLAC National Accelerator Laboratory, Menlo Park, CA 94025, USA}
\author{John Power}
\affiliation{Argonne National Laboratory, Lemont, IL 60439, USA}


\begin{abstract}
Beams with cross-plane coupling or extreme asymmetries between the two transverse phase spaces are often encountered in particle accelerators. Flat beams with large transverse-emittance ratios are critical for future linear colliders. Similarly, magnetized beams with significant cross-plane coupling are expected to enhance the performance of electron cooling in hadron beams. Preparing these beams requires precise control and characterization of the four-dimensional transverse phase space. In this study, we employ generative phase space reconstruction techniques to rapidly characterize magnetized and flat-beam phase-space distributions using a conventional quadrupole-scan method. The reconstruction technique is experimentally demonstrated on an electron beam produced at the Argonne Wakefield Accelerator and  successfully benchmarked against conventional diagnostics techniques. 
Specifically, we show that predicted beam parameters from the reconstructed phase-space distributions  (e.g. as magnetization and flat beam emittances) are in excellent agreement with those measured from the conventional diagnostic methods. 
\end{abstract}

\maketitle

\section{Introduction}

Using magnetized electron beams, whose transverse motion is dominated by angular momentum, to cool hadron beams in circular accelerators has been proposed as an effective way to reduce the transverse emittance of beams in colliders ~\cite{Budker:1967sd, Budker:1021068, Derbenev:1978, Derbenev:1978qd}.
In this process, a magnetized electron beam co-propagates with a hadron beam in the presence of a solenoid magnetic field, transferring thermal energy from the hadron beam to the electrons, thus improving the emittance of the hadron beam.
Generating magnetized electron beams for this purpose using non-zero solenoid fields on the cathode of electron photoinjectors is a promising area of active study~\cite{PhysRevSTAB.7.123501, WIJETHUNGA2023168194}.

Furthermore, it is critical to preserve the beam magnetization during transport between the electron source and the electron-hadron interaction point.
To overcome the issue of the magnetized beam propagation for the case where the beam energy is large ($\gamma>50$), the use of skew quadrupole magnets to eliminate transverse coupling was proposed and demonstrated, enabling the use of existing normal quadrupole magnets to transport the decoupled beam~\cite{PhysRevSTAB.4.053501, PhysRevSTAB.9.031001, Halavanau:2018hcg, Xu:2019xjg, FETTERMAN2022166051}. 
Afterward, the decoupled electron beam is converted back to magnetized state again using the skew quadrupole magnets for the actual application. Here, during the magnetized beam transformation, the beam phase space is re-partitioned to each term of particle momentum in the horizontal and vertical planes. This leads to asymmetry of electron beam emittance in the transverse plane~\cite{PhysRevSTAB.4.053501}.
This beam is called a flat-beam, where the large emittance ratio is associated with the magnetization.
Flat beam distributions are not only applicable in the context of preserving magnetization but also useful for collider applications to enhance the luminosity at the interaction point~\cite{White_2022, PhysRevAccelBeams.26.014001}.

Finally, developing effective beam diagnostics to precisely characterize the magnetization of electron beams is an equally important task required to match the beam dynamics during electron-hadron co-propagation in the solenoid magnet; namely maintaining a constant electron beam envelope as it travels through the solenoid magnetic field~\cite{wang2003simulation, kewisch2004magnetized}.
Accordingly, beam diagnostics are required to characterize (1) the magnitude of beam magnetization before the flat beam transformation, (2) the large ratio of transverse beam emittances, and (3) a lack of coupling between horizontal and vertical phase spaces after the flat beam transformation. 
Measurements of the beam magnetization have previously been conducted using a collimating slit method~\cite{PhysRevAccelBeams.22.102801, PhysRevAccelBeams.25.044001} which estimates beam magnetization by measuring the transverse angular deflection of the beam after propagating through a drift. 
This method requires specialized diagnostic elements (motorized slit) and only measures the magnetization of a subset of the beam distribution, leading to over- or under-estimation of the magnetization of the full beam.

\begin{figure*}
   \centering
   \includegraphics[width=\textwidth]{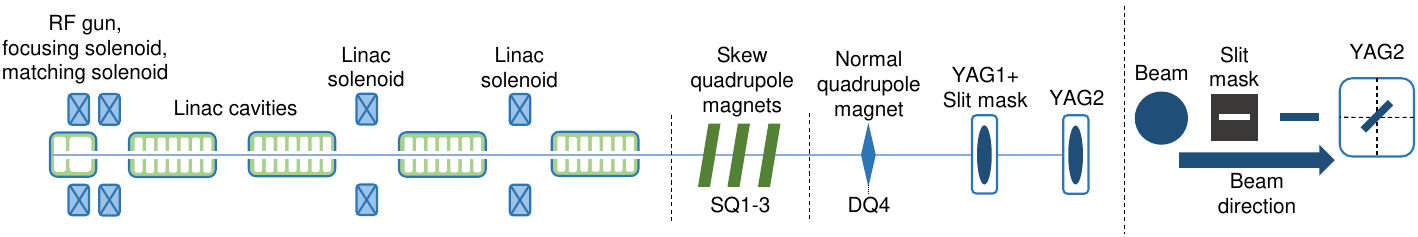}
   \caption{AWA drive linac and beam transport line used in this demonstration (not to scale) \cite{kim:ipac2023-wepa037}. 
   On the right, a cartoon is introduced on measuring the magnetization $\Lb$ using slit and YAG screens.}
   \label{fig:AWASchematic}
\end{figure*}

Quadrupole scans using a single magnet are commonly used to measure the transverse beam emittances. 
However, they only provides scalar estimates of the beam distribution (as opposed to detailed information) and does not characterize the transverse coupling between the horizontal and vertical phase spaces. 
Characterizing the transverse coupling requires additional measurements of the beam propagating through a long drift section which can be time-intensive and require specific beamline configurations.
The practical use of flat beams in accelerator applications requires fast methods for characterizing the four-dimensional transverse phase space with and without coupling terms between the horizontal and vertical phase spaces.

To solve this problem, we utilize a recently developed method for reconstructing detailed phase space distributions from screen images using generative machine learning (ML) models ~\cite{PhysRevLett.130.145001}, referred to here as generative phase space reconstruction (GPSR). 
The GPSR method has two key concepts in order to reconstruct beam distributions: (1) generative ML models that can parameterize arbitrary beam distributions in phase space with unprecedented levels of detail and complexity, and (2) \emph{differentiable} beam dynamics simulations that enable inexpensive calculations of model output gradients \cite{gonzalez-aguilera:ipac2023-wepa065} which are used to train the ML-based representation of the beam distribution from experimental data. 
Initial demonstration~\cite{PhysRevLett.130.145001} shows that this method can successfully reconstruct four-dimensional phase space distributions of beams from basic quadrupole scan measurements in both simulation and experiment. 

In this paper, we apply GPSR to reconstructing magnetized and flat beam distributions using quadrupole scan measurements.
We first conduct a reconstruction of a magnetized beam distribution from experimental data measured at the Argonne Wakefield Accelerator (AWA) and compare estimates of the beam magnetization from the reconstruction with conventional diagnostic methods.
We then examine the application of GPSR to reconstructing a flat beam distribution to confirm a large beam emittance ratio between the horizontal and vertical transverse phase spaces.

In the following section, we introduce the experimental settings performed at the AWA facility.
In Sec.~\ref{sec:magbeam_reconstruction}, we demonstrate phase space reconstruction and comparison of beam parameters in case of magnetized beam.
Then, in Sec.~\ref{sec:flatbeam_reconstruction}, we cover the discussion of reconstructed phase space for a transformed flat beam.
Finally, we conclude with a discussion of the results and propose future work in Sec.~\ref{conclusion}.
Details of the phase space reconstruction algorithm, data acquisition, and discussions on uncertainty of measurements are described in Appendices~\ref{appenx_gpsr} and \ref{appenx_uncertainty}.

\section{Experimental setup and brief fundamentals of flat and magnetized beams}

Figure~\ref{fig:AWASchematic} presents a schematic of the AWA beamline. 
The beamline consists of an RF gun, several solenoid magnets, and linear accelerating cavities.
The solenoid magnet nominally used to buck the magnetic field on the photocathode was used during this work to generate a non-zero magnetic field at the cathode surface to produce a transversely coupled beam distribution that has a non-zero angular momentum. 
The nominal beam charge used for this demonstration is 1.0~nC.
The measured beam energy using the spectrometer magnet at the end of the beamline is approximately 43.4 MeV.
This AWA beamline also includes three skew quadrupole magnets (SQ1-3) and a normal quadrupole magnet (DQ4).
The DQ4 magnet is used in this work to rotate the transverse phase space of the distribution (often referred to as a quadrupole scan) which is then imaged on a downstream YAG screen (YAG1).
The magnetic length of the DQ4 quadrupole magnet is 0.12~m, and the distance between DQ4 and YAG1 is about 1.33 m center-to-center.
An additional YAG screen, indicated as YAG2 on Fig.~\ref{fig:AWASchematic}, is used to estimate the magnetization by measuring the RMS beam size and capturing the angle of the beam slice when the beam is cut by the slit at the upstream YAG1 position.\\

\begin{figure*}
   \centering
   \includegraphics[width=\textwidth, keepaspectratio]{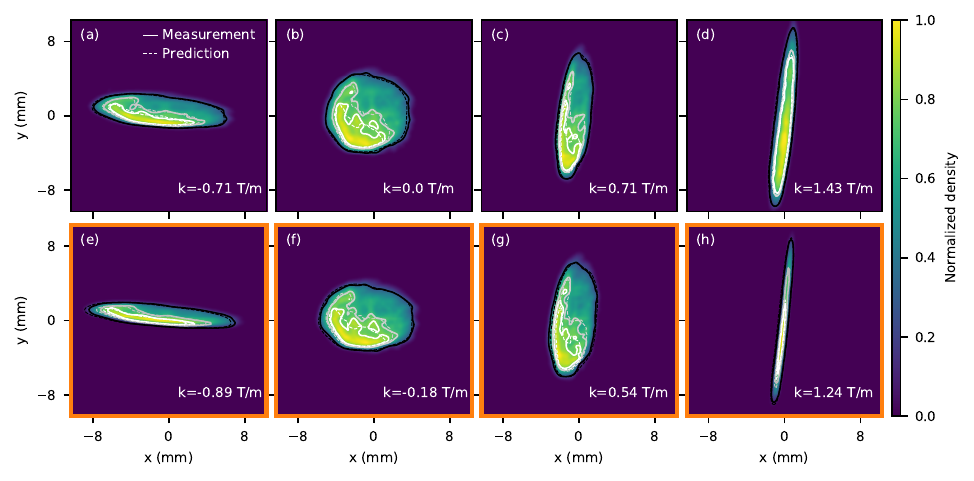}
   \caption{Measured magnetized beam images at YAG1 screen and contour lines with different quadrupole strength. Black-gray-white contour lines represent $95-50-25$ percentile of the beam distribution. Solid lines indicate the measured distribution, and dashed lines show the predictions from the reconstructed beam. (a-d) show the training dataset, while (e-h) with orange border line show the test set. Quadrupole magnet strengths are indicated on the bottom right of each sub-figure. }
   \label{fig:magnetized_traintest}
\end{figure*}

Throughout this paper we consider the particle motion in the four-dimensional phase space $(x,x',y,y')$ where $(x,y)$ is the position coordinate and $(x',y')$ refers to the angle with respect to the direction of motion $\hat{z}$. To describe the magnetized beam, we start from the canonical momentum $\mathbf{P}$ of an electron experiencing a magnetic field at the photocathode surface 
\begin{equation}
\mathbf{P}=\mathbf{p}+e\mathbf{A},
\label{eq:canonicalangularmomentum}
\end{equation}
where $\mathbf{p}$ is the kinetic momentum and $\mathbf{A}=\frac{B_{c}}{2}(-y\hat x + x\hat y)$ is the vector potential produced by the solenoid with magnetic-field value $B_{c}$ at the photocathode surface. Note that  $\mathbf{A}=A\hat{\theta}$ and $|e\mathbf{A}| \gg |\mathbf{p}|$ during the emission process so that the beam is born with a non-vanishing canonical angular momentum. Upon exiting the solenoid, the conservation of angular momentum yields the kinetic momentum to acquire a large angular component resulting in an angular-momentum-dominated beam.
Correspondingly, the canonical angular momentum $\mathbf{L}$ becomes
\begin{equation}
\textbf{L} = \mathbf{r} \times \mathbf{P} = (x\hat x + y\hat y)\times e\mathbf{A} = \frac{eB_{c}}{2}(x^{2}+y^{2})\hat z,
\label{eq:angularmomentum}
\end{equation}
The ensemble-averaged angular momentum is called the magnetization and is given by
\begin{equation}
\Lb = \frac{\braket{|\textbf{L}|}}{2m_{e}c} = \frac{eB_{c}\sigma_{c}^{2}}{2m_{e}c},
\label{eq:magnetizationo}
\end{equation}
where $m_{e}$ is the electron mass, and $c$ is the speed of light, $e$ is an elementary charge, and $\sigma_{c}$ is the RMS transverse spot size of the beam distribution on the cathode, respectively. The $\braket{\ldots}$ indicates the statistical averaging over the distribution.

The four-dimensional beam matrix associated with the angular-momentum-dominated beam can be written as~\cite{Kim:PhysRevSTAB.6.104002} 
\begin{equation}
\Sigma = \left[\begin{array}{cc} 
\epsilon_{eff}T_{x} & \Lb J \\
-\Lb J & \epsilon_{eff}T_{y} \\
\end{array}\right],
\label{eq:sigmamatrixb}
\end{equation}
where 
\begin{equation}
T_{x,y} = \left[\begin{array}{cc} 
\beta_{x,y} & -\alpha_{x,y} \\
-\alpha_{x,y} & \frac{1+\alpha_{x,y}^{2}}{\beta_{x,y}^{2}} \\
\end{array}\right],~~ J = \left[\begin{array}{rr} 
0 & 1 \\
-1 & 0 \\
\end{array}\right].
\label{eq:Tmatrix}
\end{equation}
$\beta_{x,y}$ and $\alpha_{x,y}$ are Twiss parameters of the beam.
Here, $\epsilon_{eff}=\sqrt{\epsilon_{u}^{2} + \Lb^{2}}$ where $\epsilon_{u}$ is the uncorrelated emittance.

Using a proper symplectic transformation, the cross-plane coupling can be removed resulting in a final beam matrix that is $2\times2$ block diagonal as described in Eq.\eqref{eq:sigmamatrix_diagonalized}.
\begin{equation}
\Sigma_{f} = M\Sigma M^{T} = \left[\begin{array}{cc} 
\epsilon_{+}T_{+} & 0 \\
0 & \epsilon_{-}T_{-} \\
\end{array}\right], 
\label{eq:sigmamatrix_diagonalized}
\end{equation}
where $\epsilon_{\pm}$ are the eigenemittances, which can be obtained from the feature of the invariance of 4D emittance $\epsilon_{4D}=\sqrt{det(\Sigma})$~\cite{PhysRevLett.64.1073}
\begin{equation}
\epsilon_{\pm} = \sqrt{\epsilon_{u}^{2} + \Lb^{2}} \pm \Lb.
\label{eq:eigenemittances}
\end{equation}

The process of diagonalizing the four-dimensional beam matrix is practically implemented using three skew quadrupole magnets~\cite{PhysRevSTAB.4.053501} -- a beamline often referred to as round-to-flat-beam transformer (RFBT). 
As a result, the diagonalized beam has eigenemittances that correspond to the projected transverse emittances.
When the magnetization satisfies $\Lb \gg \epsilon_{u}$, the eigenemittances simplify as 
\begin{equation}
\epsilon_{+} \simeq 2\Lb, ~~ \epsilon_{-} \simeq \frac{\epsilon_{u}^{2}}{2\Lb}.
\label{eq:eigenemittances2}
\end{equation}
Thus, as $\Lb$ increases by applying a stronger magnetic field at the cathode, the emittance of the flat beam becomes increasingly  asymmetric.
Detailed descriptions of the optimal skew quadrupole strengths for achieving the RFBT condition (and vice-versa), given the incoming beam parameters, are detailed in Refs.~\cite{thrane2002photoinjector, PhysRevSTAB.9.024001}.

\begin{figure}
   \centering
   \includegraphics[width=8.5 cm, keepaspectratio]{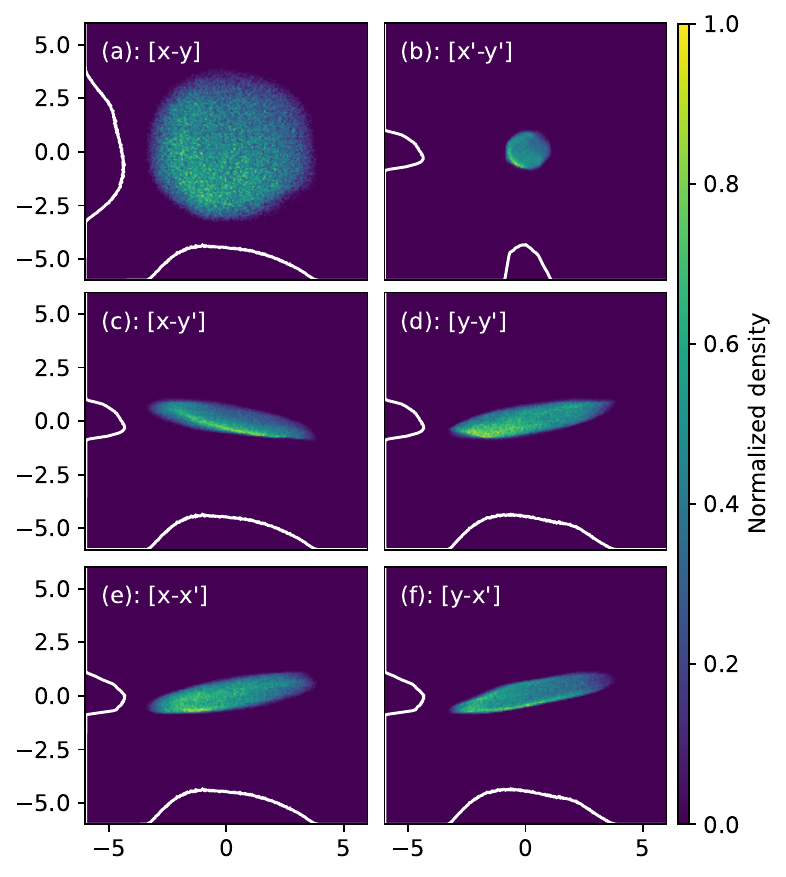}
   \caption{Reconstructed phase space of the magnetized beam in front of DQ4 quadrupole magnet. $x'$ and $y'$ are normalized quantities by the reference momentum that corresponds to the kinetic energy of 43.4~MeV.
   Here, the center of the phase space is shifted by considering its original mean position.
   White lines denote density projections onto the horizontal and vertical axes. 
   All units in the sub-figure are mm or mrad.}
   \label{fig:magnetizedbeam_recon}
\end{figure}

\section{Phase space reconstruction for a magnetized beam}\label{sec:magbeam_reconstruction}
First, we will discuss the phase space reconstruction for the magnetized beam case.
With the skew and normal focusing quadrupoles turned off, the magnetized beam was transported to and imaged on YAG1.
The normal quadrupole magnet (DQ4) gradient was scanned from $-1.79$~T/m to $1.79$~T/m over 21 steps with 3 images captured at each quadrupole-magnet settings.
A subset of the images captured appear in Fig.~\ref{fig:magnetized_traintest}.
Half of the measured cases were then passed to the GPSR algorithm to reconstruct the beam distribution, with the other half of the cases set aside to be used as a test set to validate the reconstruction accuracy.
Details regarding the image processing and the reconstruction algorithm are described in the Appendix~\ref{appenx_gpsr}.

Black, gray, and white lines on Fig.~\ref{fig:magnetized_traintest} indicate 95th, 50th, and 25th percentile contour lines of the measured and predicted beam densities for each quadrupole strength respectively.
In the case of the measured beam density, high-frequency modulations are observed around the core of the measured beam density.
These features originate
from the micro-lens array (MLA)~\cite{PhysRevAccelBeams.20.103404} nominally used to homogenize the transverse distribution of the UV laser pulse on the photocathode surface. 
The MLA yields highfrequency modulations arising from the formation of local focal points from the microlenses composing the array.
Nevertheless, we observe that the predicted density distribution closely matches the characteristics of the training data.
Furthermore, predictions of the beam distribution as a function of normal quadrupole focusing strength also agree well with the test measurements excluded from the training set.
As a result, we conclude that the reconstructed transverse beam distribution matches the real beam distribution.

The reconstructed phase space at the entrance of the DQ4 quadrupole magnet is illustrated in Fig.~\ref{fig:magnetizedbeam_recon}.
The $(x-y)$ distribution shown in Fig.~\ref{fig:magnetizedbeam_recon}(a) implies that the beam is nearly round with beam sizes of $\sigma_{x}(\sigma_{y})=1.62(1.58)$~mm and projected emittance $\epsilon_{nx}(\epsilon_{ny})=48.6(44.9)$~mm~mrad, respectively, where the emittance has the contribution from the canonical angular momentum terms.
As can be seen in Fig.~\ref{fig:magnetizedbeam_recon}(d,e), the position-divergence correlations within the projected  phase spaces $(x-x')$ and $(y-y')$ are almost identical to each other.

Likewise, examining the cross-plane coupling by analysing the projected phase-space distributions ($x,y'$) and ($y,x'$) confirm, albeit for some imperfections, that average coupling between the transverse positions and momenta $\gamma\braket{xy'} = -43.4$~\si{\micro\meter}  and $\gamma\braket{yx'} = 44.6$~\si{\micro\meter} are nearly equal in magnitude but opposite in sign; see  Fig.~\ref{fig:magnetizedbeam_recon}(c,f). 
This is consistent with the analytical beam matrix that describes a magnetized beam in Eq.\eqref{eq:sigmamatrixb}.
These coupling terms yield a prediction of the beam magnetization, $|\Lb|=|\gamma\left(\braket{xy'}-\braket{yx'}\right)|/2=44.0$~\si{\micro\meter}.
Compared to the measured $\Lb$ of $47.2\pm0.9$~\si{\micro\meter} that was obtained by using the slit~\cite{PhysRevAccelBeams.22.102801} and measuring the angle of the beam slice (see Fig.~\ref{fig:AWASchematic}), the $\Lb$ from the reconstruction shows good agreement within 10\% error. 
Here, the error of the measured $\Lb$ was determined by considering the standard deviation of the RMS beam size and angle of the beam slice where the angle was determined by using linear fitting.

\begin{figure*}[ht]
   \centering
   \includegraphics[width=\textwidth, keepaspectratio]{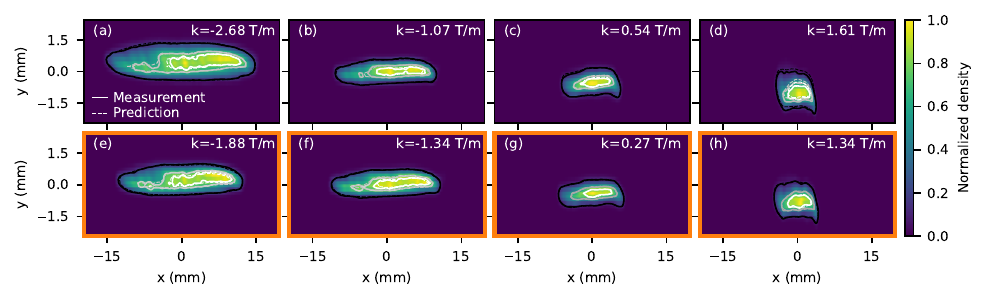}
   \caption{Measured flat beam images at YAG1 screen with different quadrupole strength. Black-gray-white contour line represents $95-50-25$ percentile of the beam distribution. Solid lines indicates the contour line for measured distribution, and dashed lines show the contour lines for the reconstructed beam. (a-d) show train dataset, while (e-h) with orange border line show test set. On top right of subfigure, strength of the quadrupole magnet is indicated.}
   \label{fig:flatbeam_traintest}
\end{figure*}

\section{Phase space reconstruction for a flat beam}\label{sec:flatbeam_reconstruction}
We then demonstrate phase space reconstruction of a flat beam distribution to estimate the ratio of transverse beam emittances after the RFBT.
The skew quadruople magnet focusing strengths were adjusted such that they applied the RFBT to the magnetized beam, removing the cross-plane correlation and partitioning the beam emittances along the principal horizontal and vertical axes.
Once the transformation was established, we repeated the quadrupole scan performed in Sec. \ref{sec:magbeam_reconstruction}.
Here, note that the applied solenoid field at the cathode for this flat beam case is different compared to that in Sec.~\ref{sec:magbeam_reconstruction}.
The magnetic field gradient of the quadupole magnet was adjusted from $-3.2$~T/m to $2.4$~T/m over 22 steps to cover the entire evolution of the beam focusing in horizontal and vertical planes.
Likewise, we acquired 3 images for each quadrupole-magnet setting for the reconstruction process.

Figure~\ref{fig:flatbeam_traintest} shows the measured beam distribution with different quadrupole strengths at YAG1. 
Similar to Fig.~\ref{fig:magnetized_traintest}, the top row shows the training dataset, whereas the bottom row describes test data used to validate the reconstruction accuracy. 
We again observe that the contour areas predicted by the reconstruction well represents those from the measurements, including a nonlinear distortion in the beam distribution in Fig.~\ref{fig:flatbeam_traintest}(d,h) as the beam is strongly focused in the horizontal plane. 
We hypothesize that the main source of this distortion is the residual nonlinear space charge effect associated with the transverse beam distribution which is slightly deviated from elliptical shape~\cite{PhysRevAccelBeams.25.044001}.

\begin{figure}[t]
   \centering
   \includegraphics[width=8.5 cm, keepaspectratio]{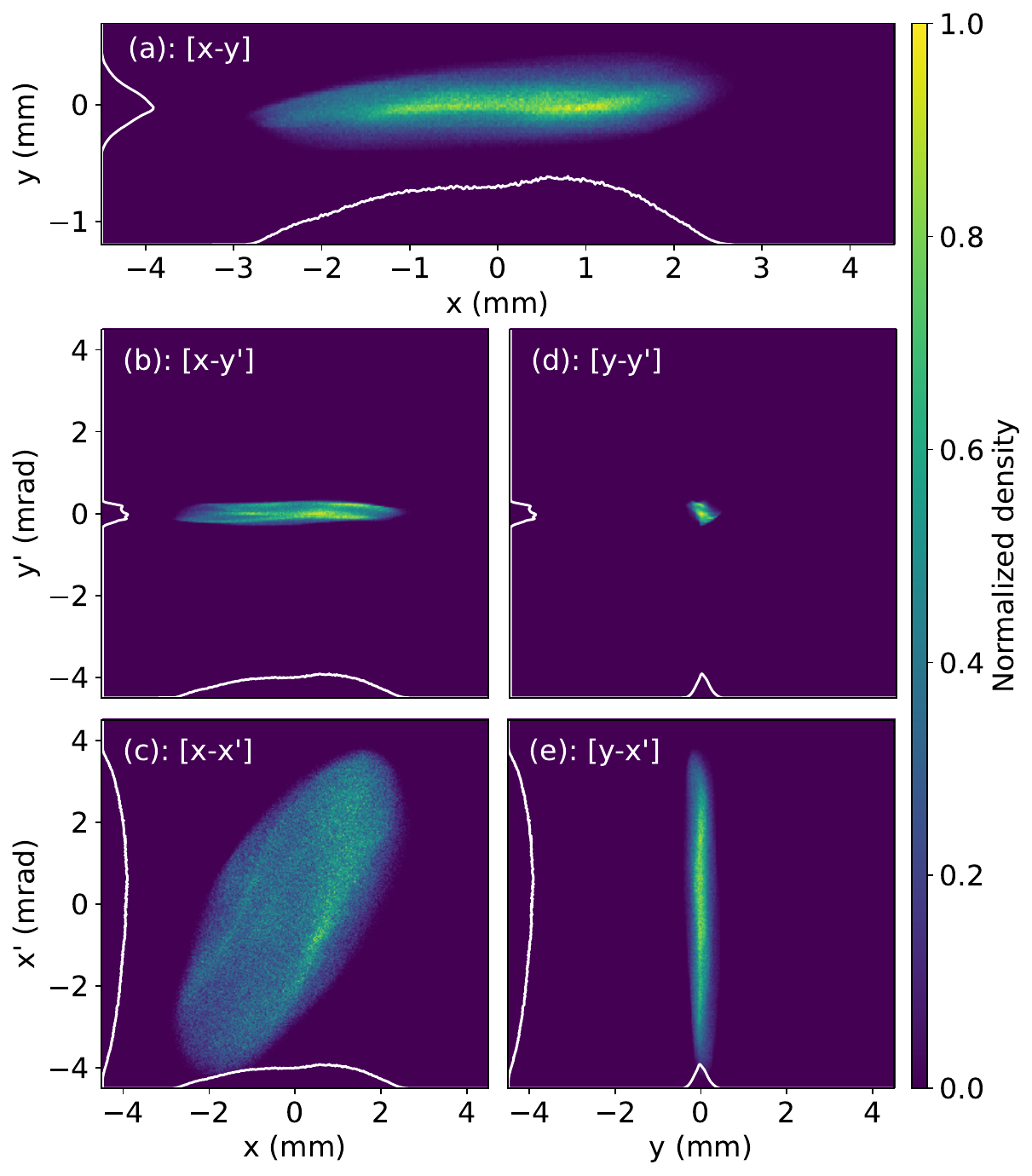}
   \caption{Reconstructed phase space of the flat beam in front of DQ4 quadrupole magnet.}
   \label{fig:flatbeam_recon}
\end{figure}

The reconstructed 4D phase space at the entrance of DQ4 quadrupole magnet appears in Fig.~\ref{fig:flatbeam_recon}.
Figure~\ref{fig:flatbeam_recon}(b,e) qualitatively show that the transverse correlations are minimized, consistent with the beam matrix given by Eq.\eqref{eq:sigmamatrix_diagonalized} when the angular momentum of the beam is minimized.
The reconstruction predicts that the magnetization $\Lb$ of the flat beam is $3.1$~\si{\micro\meter}, an order of magnitude smaller than that of the round beam before the round to flat beam transformation. 
Additionally, as described in Appendix 3, correlations between the horizontal and vertical phase space coordinates are minimized.

In addition, we can see the large emittance ratio between the horizontal and vertical phase spaces, shown in Fig.~\ref{fig:flatbeam_recon}(a,c,d).
This is the expected feature from Eqs.(\ref{eq:sigmamatrix_diagonalized}, \ref{eq:eigenemittances}) where the eigenemittances in the horizontal and vertical planes are largely dominated by the magnetization, and those terms will contribute to the block diagonal terms of the beam matrix.
Therefore, it is expected that the flatness of the beam is maintained.

\begin{table}[t]
\caption{
Comparison of the flat-beam emittance and Twiss parameters. Measurement error source is the standard deviation of the measured beam size.
}
\begin{ruledtabular}
\begin{tabular}{l r r r}
\textrm{Case} & \textrm{Measurement} & \textrm{Reconstruction} & \textrm{unit}\\
\colrule
$\epsilon_{nx}$ & 144.64$\pm$1.36 & 140.14 & mm~mrad\\
$\epsilon_{ny}$ & 1.47$\pm$0.10 & 1.53 & mm~mrad\\
$\beta_{x}$ & 0.93$\pm$0.02 & 0.88 & m\\
$\beta_{y}$ & 1.42$\pm$0.04 & 1.14 & m\\
$\alpha_{x}$ & $-1.02\pm0.02$ & $-0.90$ & rad\\
$\alpha_{y}$ & 1.61$\pm0.02$ & 0.43 & rad\\
\end{tabular}
\end{ruledtabular}
\label{table:flatbeamemittance} 
\end{table}

We also compared scalar characterizations of the reconstructed transverse phase space with scalar measurements from quadrupole scan in Table~\ref{table:flatbeamemittance}.
We see good agreement ($<10$\%) between most of the scalar parameters determined by the quadrupole scan measurements with the reconstructed beam distribution, with the exception of the vertical Twiss parameters.

A likely cause of the discrepancy between the vertical Twiss parameters is the limited vertical resolution with the phase space reconstruction algorithm. 
The large aspect ratio of the flat-beam distribution makes it difficult to accurately resolve changes in the vertical beam size as a function of quadrupole strength.
Using a diagnostic with a higher resolution or a beam manipulation that can increase the phase advance of the vertical phase space would likely remedy this issue.
Alternatively, the GPSR algorithm can be used to stitch together multiple, magnified images of individual portions of the beam distribution, resulting in a higher resolution multi-shot measurement.

Finally, the original magnetization can be estimated from the reconstructed flat beam.
According to Eq.\eqref{eq:eigenemittances2}, the horizontal emittance is approximately 2 times larger than the $\Lb$.
Therefore, the resultant $\Lb$ from the reconstruction becomes 70.1~\si{\micro\meter}.
We confirmed that the reconstructed beam magnetization has quantitative agreement within 10\% error compared to the measured value $\Lb=59.3\pm 7.3$~\si{\micro\meter} when considering $1\sigma$ of the measured magnetization, where the measurement was done using the magnetized beam before the RFBT and method introduced in Sec.~\ref{sec:magbeam_reconstruction}.
The source of the measurement error is same as what is described in the previous section.

\section{Conclusion}\label{conclusion}

In this work, we have demonstrated that phase space reconstruction using the GPSR technique is effective at accurately reconstructing magnetized and flat beam distributions at the AWA using simple quadrupole scans, as well as accurately estimating key parameters of interest, including the beam magnetization and emittance ratios.
Particularly, this phase space reconstruction technique can capture the entire information of four-dimensional beam phase space without additional beam diagnostic tool such as the slit to measure the cross-plane correlations.

To enhance the performance of the GPSR technique, such as capturing higher frequency features due to the use of the MLA, we will perform further investigation such as the use of increased number of particles during the reconstruction and kernel density estimation with different method in addition to the Gaussian distribution~\cite{wkeglarczyk2018kernel, KRISTAN20112630}.
In addition, including space charge effect during the GPSR process is considered as one of the future works.

In addition to reconstructing the four-dimensional phase space, the demonstrated technique could be further extended by adding a transverse-deflecting cavity and dipole magnet to enable the reconstruction of the complete six-dimensional phase space using the GPSR method.
Therefore, this will opens a path for data-based, physics-informed reconstruction for experiments that require the understanding of transverse-longitudinal couplings along emittance exchange (EEX)~\cite{PhysRevLett.118.104801} and double-emittance exchange (DEEX)~\cite{PhysRevLett.129.224801} beamlines for 6D beam phase-space manipulations.

\begin{acknowledgments}
This work was supported by the U.S. Department of Energy award No. DE-AC02-06CH11357 to Argonne National Laboratory and by the U.S. National Science Foundation under the award PHY-1549132, the Center for Bright Beams. This work was also funded by the U.S. Department of Energy, Office of Science, Office of Basic Energy Sciences under Contract No. DE-AC02-76SF00515. 
This research used resources of the National Energy Research Scientific Computing Center (NERSC), a U.S. Department of Energy Office of Science User Facility located at Lawrence Berkeley National Laboratory, operated under Contract No. DE-AC02-05CH11231 using NERSC award BES-ERCAP0020725.
\end{acknowledgments}

\appendix
\section{Data acquisition and GPSR setup}\label{appenx_gpsr}
First, we obtained the measured dataset for flat and magnetized beams. 
In the magnetized beam case, first we took 15 images at each quadrupole strength by applying a charge window of $\pm3$\%.
Among them, we used only three samples that are close to the nominal charge of 1.0~nC.
Likewise, in case of the flat beam, we measured 10 shots with a charge window of $\pm5$\% from nominal 1.0~nC for each case of the quadrupole magnet strength.
Afterwards, we again took only three samples with charge levels close to the nominal value.

The initial number of pixels of the screen image is 1000$\times$1000 both for flat and magnetized beams, but it was down-scaled to 250$\times$250 due to the memory resources of the computation.
Before down-scaling the image, we subtract a threshold value to remove the background noise; the value obtained by the triangle threshold method~\cite{zack-1977-a} was scaled by multiplying by 1.7.
In addition, a Gaussian filter with a standard deviation of one pixel was applied on the beam image.

\begin{table}[t]
\caption{
Parameters that are used in the phase space reconstruction.
}
\begin{ruledtabular}
\begin{tabular}{l r }
\textrm{Parameters} & \textrm{Value} \\
\colrule
Number of reconstructed particles & 100,000 \\
Number of epochs & 4,000\\
Number of neural network layers & 2 \\
Number of neurons in each layer & 20\\
Loss parameter $\lambda$ & $10^{10}$--$10^{13}$ \\
\end{tabular}
\end{ruledtabular}
\label{table:reconparametertable} 
\end{table}

For the phase space reconstruction process, we used the {\sc Python} package {\sc Pytorch}~\cite{NEURIPS2019_bdbca288}. 
For the snapshot ensembling, we utilized {\sc TorchEnsemble}~\cite{torchensemble}.
Table~\ref{table:reconparametertable} shows the parameter lists that are used for the phase space reconstruction.
The number of epochs was set to 4,000. 
The number of reconstructed particles was set to $10^{5}$ to ensure that the reconstruction successfully predicts the feature of the beam distributions.
Within these epochs, we clearly saw that the transverse 4D parameters are well converged even under snapshot ensembling to check whether the reconstructed result is not stuck in local minima.
Here, the minimum and maximum learning rates were $10^{-5}$ and $10^{1}$, respectively.
The loss parameter, $\lambda$, which is the parameter for evaluating the loss associated with the beam's six-dimensional emittance (stated as entropy)~\cite{PhysRevLett.130.145001} is varied from $10^{10}$ to $10^{13}$, and we observed that the reconstruction result converged.
The reconstruction process was performed using the {\sc Perlmutter} cluster at NERSC, and its execution time was about several minutes.
Details of the GPSR method can be found in the supplemental material of Ref.~\cite{PhysRevLett.130.145001}.

\section{Uncertainty of Measurements}\label{appenx_uncertainty}
The radius of the UV laser at the virtual cathode is set to approximately 2.7~mm during the experiment.
In the case of the transverse distribution of the UV laser, we tried to make the distribution uniform using the MLA.
For the case described in Sec.~\ref{sec:magbeam_reconstruction}, the solenoid field at the cathode is set to 0.1~T.
Thus, the theoretically calculated $\Lb$ is 53.4~\si{\micro\meter}.
Likewise, the solenoid field used for the case in Sec.~\ref{sec:flatbeam_reconstruction} is 0.14~T, which corresponds to $\Lb=74.7~\si{\micro\meter}$.
The measured values are slightly deviated from those theoretical estimations. 
It is expected that the measured $\Lb$ using the YAG screens and slit has some systematic errors in addition to the measurement jitter, such as setting of the camera gain level that determines the amount of captured light of the scintillation screen (including the virtual cathode image).
Particularly, the RMS quantity strongly affects the measured $\Lb$ using the slit; when there is 3\% error of the RMS beam size at each YAG screen, then these will contribute to the total error of 6\% on the $\Lb$ of our reconstruction cases.
Moreover, the angle of the beam slice may have an error range of $\pm1$ degree due to the non-linearity of the slice. This will lead to $\pm[6-8]$\% changes on the $\Lb$ of the demonstrated cases.
Therefore, under consideration of the uncertainty in the measurements, the reconstructed beam phase space well predicts the transverse emittance and cross-plane correlations of the measured beam distributions.

\section{Evaluation of Correlations}

In order to discuss the transverse coupling quantitatively, we computed the Pearson correlation coefficient~\cite{Benesty2009} on the prediction of the reconstructed flat and magnetized beam phase spaces.
A correlation coefficient close to zero indicates a weak linear correlation between variables, and a correlation coefficient close to $\pm 1$ indicates strong linear correlation.

In the reconstructed beam cases, the correlation coefficients are shown in Eqs.(\ref{eq:correlation_coe_mag}, \ref{eq:correlation_coe_flat}), where $\rho_{mag}$ shows the coefficients for magnetized beam and $\rho_{flat}$ for flat beam, respectively.
The correlations were calculated under the transverse coordinate system; transverse coordinate system; $\rho(i,j)=<X_i X_j>/<X_i^2>^{1/2}< X_j^2>^{1/2}$ where $X=(x,x',y,y')$. 
\begin{equation}
\rho_{mag} = \left[\begin{array}{rrrr} 
1.000 & 0.644 & -0.013 & -0.705 \\
0.644 & 1.000 & 0.719 & 0.019 \\
-0.013 & 0.719 & 1.000 & 0.664 \\
-0.705 & 0.019 & 0.664 & 1.000 \\
\end{array}\right],
\label{eq:correlation_coe_mag}
\end{equation}
\begin{equation}
\rho_{flat} = \left[\begin{array}{rrrr} 
1.000 & 0.673 & 0.295 & 0.189 \\
0.673 & 1.000 & -0.158 & 0.246 \\
0.295 & -0.158 & 1.000 & -0.402 \\
0.189 & 0.246 & -0.402 & 1.000 \\
\end{array}\right].
\label{eq:correlation_coe_flat}
\end{equation}

The transverse coupling between $x[y]$ and $y'[x']$ can be estimated by $\rho(1,4)[\rho(3,2)]$.
As can be seen, the magnitude of the the correlation coefficients of the magnetized beam becomes close to 1, while it is close to zero for the flat beam case.
In case of the transformed flat beam obtained by the AWA drive linac simulation, magnitude of the coefficients regarding $(x-y)$ and $(x'-y')$ is similar to that in Eqs.\eqref{eq:correlation_coe_flat}.
We found that the space charge effect affects the transformed flat beam, causing nonlinear distortions in $(x-y)$ and $(x'-y')$ phase spaces that lead to non-zero coefficients. 
Thus, it is expected that the space charge effect is inherent in the measured flat beam through the prediction using the GPSR method.

%

\end{document}